\documentclass[a4paper,12pt]{article}
\usepackage{cite}
\usepackage{amssymb,amsmath}
\newcommand{\be}{\begin{equation}}
\newcommand{\ee}{\end{equation}}
\newcommand{\bea}{\begin{eqnarray}}

\newcommand{\eea}{\end{eqnarray}}

\begin{document}

\title{A Brief Editorial on de Sitter Radiation via Tunneling}

\author{A.J.M. Medved \\ \\
Physics Department \\
University of Seoul \\
 Seoul 130-743\\
Korea \\
E-Mail(1): allan@physics.uos.ac.kr \\ 
E-Mail(2):  joey\_medved@yahoo.com \\ \\}

\maketitle
\begin{abstract}

Very recently (by which we mean two days ago),
Y. Sekiwa submitted a paper \cite{sekiwa} that dismisses some
previous calculations of the de Sitter radiative temperature.
Significantly, these calculations employed 
the (so-called) tunneling methodology, which  
the stated author then goes on to correspondingly revise. After  briefly commenting
on  what the tunneling framework can  be claimed to say  and what it cannot, 
we respond to the author's criticisms and  proposed revision.

\end{abstract}
\newpage
\section*{}

The tunneling method of Parikh and Wilczek (PW) \cite{pw} has proven 
to
be a useful analytic tool for computing the Hawking temperature \cite{haw} 
 for a wide assortment of black
hole (and related) models. (This includes confirmations of
the tree-level value  of the temperature, as well as   more speculative corrections
that should arise due to back-reaction effects \cite{kw}.)~\footnote{See the article of
our particular interest \cite{sekiwa} for many 
applicable references or, alternatively, consult your local search 
engine.} The PW  framework follows from the intuitively
simple and enticingly picturesque notion of the Hawking radiative  process as
 a  (quantum) pair-production event.  To elaborate on this premise, a pair of particles is  
induced --- quantum gravitationally ---  into existence just inside of
the  horizon of a black hole. This process
is  followed by the positive-energy particle quantum-tunneling its
way through the horizon. Ultimately, such particles will escape from the 
near-horizon region to become the observable quanta of Hawking radiation,
while their negative-energy partners remain behind to lower the mass
of the black hole.

In spite of its obvious attributes (or maybe because of
these), the tunneling framework has come under some recent scrutiny,
which has turned up some disturbing inconsistencies.
The most glaring of which is a lack of invariance of
the formalism when subjected to conformal transformations
\cite{chow,sing}. This observation, as well as some other
calculational subtleties ({\it e.g.}, \cite{xxx-1}),
has lead to the claim that any  tunneling-model prediction
could be off by a factor of two from its Hawking-prescribed value.
If this discrepancy of two is indeed true, then what
is one to make of such an alarming development?

Let us suppose, at least for the sake of argument, that the tunneling 
interpretation 
of the Hawking radiative effect is not literally accurate. After all, 
quantum gravity is,
even at the best of times,
a rather poorly understood  regime of physics. Moreover, black
hole thermodynamics (which is best construed as a conduit of 
quantum gravity)
is,  itself,  immersed in an ethereal fog of conceptual and 
interpretational  ambiguities. Which is to say, there is  no
{\it a priori}  reason that the  Hawking effect
should necessarily have a translation that conspires to agree with
 the intuitive whims of
theoreticians. Put differently, would a classically trained physicist
of --- say --- Newton's time  have any way of  accurately casting the 
quantum
uncertainty principle  or  wave-particle duality into a physically
discernible setting?

So, if the tunneling picture of black hole radiation does turn out
to be a naive manifestation of our quantum-based biases,
where does this leave the  PW-inspired  calculations
of the Hawking temperature? As it so happens, these results can
still be 
vehemently defended;
irrespective of how one chooses to physically interpret the
underlying framework. The point here is as follows:
It has been shown --- quite conclusively by Pilling \cite{pil} (also
see \cite{yyy})  ---
that the tunneling calculation is a simple
consequence  of the first law of {\it black hole thermodynamics} \cite{haw2}
and the associated {\it area--entropy law} \cite{bek}. 
Furthermore, this observation would apply equally well to any   static and spherically symmetric
spacetime  that contains  a horizon with an analogous thermodynamic
interpretation and an areal-scaling entropy. 
Meaning that, up to the aforementioned factor-of-two discrepancy,
any such tunneling output is on firm mathematical ground; that is,
just as firm as the black hole thermodynamics that must conceptually 
underlie it. 

So, then, what about the factor-of-two dilemma? Believing the PW
formalism to be strictly a calculational tool, the tunneling skeptic
could always stick a corrective factor in by hand. However, 
 the  issue does appear to be (at least tentatively) resolved. 
This resolution was proposed by Mitra \cite{mitra} on the basis
of a previously over-looked point 
regarding  the ``principle of detailed balance".
The interested reader should consult the cited paper 
(also see \cite{paddy,vag}) for
 an elucidation.

To capsulate the above discourse, the PW tunneling formalism has
an undeniable utility as a computational tool, but it does not necessarily
provide us with anything more. Along with black hole models, this outlook should
apply equally well to de Sitter (cosmological) horizons; provided that
the  aforementioned stipulations (spherically symmetric
and static spacetime) remain in effect. Indeed,  de Sitter
horizons are well known to  have an analogous
first law and area--entropy relation, and are commonly
endowed with a similar thermodynamic  interpretation \cite{gh}.
So, it should come as no surprise that the PW formalism
has been successfully administered in just such a de Sitter context \cite{par,me}. 
 
Now we come to the crux of the matter: In a very recent
paper, Sekiwa \cite{sekiwa} has criticized these previous de Sitter 
calculations. The main point of  contention
 being that (in \cite{par,me}) the positive-energy
particle (of the quantum-produced pair) is presumed to follow 
what is a classically forbidden trajectory while escaping inward through the
de Sitter horizon.~\footnote{Opposite to  the case of a black hole,
classical matter passes through the de Sitter horizon on an outgoing
trajectory, while  quantum radiation follows an inwardly directed null geodesic.
This relative ``flip-flop" has lead to some confusion over the assignment
of signs to (de Sitter) thermodynamic quantities, but is otherwise
benign. See \cite{str} for some further discussion.} From the viewpoint
that the Hawking (or, in this case, de Sitter) radiative process
has a true tunneling interpretation, there is ---  
debatably ---
some credence to this criticism. However, 
 accepting that the PW framework is (until proven otherwise)
 really just an analytic tool, one can see that the critique
becomes quite irrelevant. To reiterate,  Pilling \cite{pil} has
shown us that the
PW formalism follows directly from the first law of
horizon mechanics and the
area--entropy law. As such, any  PW calculation is, therefore, only as credible (or not) as these
inputs; no more and no less. Until  
a {\it rigorous}  case is made for horizon radiation as a quantum-tunneling mechanism,
this is all that can be stated, definitively,  on the matter.

As for Sekiwa's proposed resolution, it is heavily founded on the idea
that the cosmological constant can be regarded as a  thermally fluctuating parameter
(see \cite{sekiwa2} and references therein).
This could very well be true; however, there is also a school of thought
that suggests otherwise. For instance, as  has  persistently  been ``preached" by 
Banks \cite{banks}, the cosmological constant
should 
not be regarded as a tuneable parameter. This
is because --- in analogy to the AdS--CFT picture of quantum gravity \cite{mal} ---
 the value of this constant determines the  entirety of the quantum-gravity  theory, 
rather than just a
particular state.

Regardless of which side of this argument one tends to favor,
it is certainly an open question as to whether the cosmological constant
should be allowed to fluctuate. Hence, Sekiwa's proposed resolution
can not be applied, consistently, to all pertaining models and interpretations
of (de Sitter) quantum gravity. Moreover, it would be unfair
to suggest (as the author goes on to do) that the tunneling framework
can be used to endorse the notion of a thermally varying
cosmological constant. Such an endorsement  would
--- by implication ---
 promote the
tunneling picture from a computational oddity 
 to a rigorous paradigm. Alas, the latter status has simply not yet
been formally realized.

Let us summarize the main doctrine of this letter: The 
FW tunneling methodology is a useful analytic
device for 
calculating the Hawking  or de Sitter temperature
of a given  gravitational model.
At the same time, the tunneling framework provides us with an intuitive  --- but
{\it non-rigorous} --- picture for the horizon-induced
radiation.  It remains an open but daunting challenge for theorists
to demonstrate if it can say or be anything more than this.

\section*{Acknowledgments}
Research is financially supported by the University of
Seoul.


\end{document}